\documentclass[prl,aps,showpacs]{revtex4}

\usepackage{epsfig}
\usepackage{amssymb}
\usepackage{bm}

\begin{document}

\title{A Future Test of Gravitation Using Galaxy Cluster Velocities}

\date{\today}
\author{Arthur Kosowsky}
\email{kosowsky@pitt.edu}
\affiliation{Department of Physics and
Astronomy, University of Pittsburgh, 3941 O'Hara Street, Pittsburgh, PA 15260 USA}
\author{Suman Bhattacharya}
\email{sumanb@lanl.gov}
\affiliation{T-2, Theoretical Division, Los Alamos National Laboratory, Los Alamos, NM 87545}

\begin{abstract}
The accelerating expansion of the Universe at recent epochs has called into question
the validity of general relativity on cosmological scales. One probe of gravity is a comparison of
expansion history of the Universe with the history of structure growth via gravitational instability: 
general relativity predicts
a specific relation between these two observables. Here we show that the mean pairwise streaming
velocity of galaxy clusters provides a useful method of constraining this relation. Galaxy cluster velocities
can be measured via the kinetic Sunyaev-Zeldovich distortion of the cosmic microwave background radiation; future surveys can provide large enough catalogs of cluster velocities to
discriminate between general relativity and other proposed gravitational theories. 

\end{abstract}


\maketitle

The most perplexing observation in physics today is the accelerating expansion of the Universe
(for a review, see \cite{fri08}). 
While such an acceleration can be brought about by a constant energy density of the vacuum,
the associated energy scale is a small
fraction of an electron volt. This energy scale is not, as far as we know, 
a natural fundamental scale in physics, and skepticism is warranted about new fundamental
physics at room-temperature energy scales which only manifests itself in cosmological phenomena. 
%(See, however, \cite{weiner,stephon} for novel attempts to connect this energy scale to fundamental physics.)

The standard hot big bang model of cosmology assumes
that the Universe is statistically homogeneous and isotropic, and that its dynamics are determined
by general relativity. The Einstein Equation describing the evolution of the metric then
reduces to the Friedmann Equation, which can be written in the form
\begin{equation}
\frac{\ddot a(t)}{a(t)} =  -\frac{4\pi G}{3}\left(\rho(t) + 3 p(t)\right)
\label{friedmann}
\end{equation}
where $\rho(t)$ and $p(t)$ are the mean energy density and pressure of the Universe at a time $t$, 
and $a(t)$ is the scale factor,
giving the ratio of the separation between two objects in the cosmic rest frame at time $t$ 
to their separation today. The scale factor is the function which describes the expansion
history of the Universe, and the Friedmann Equation is its dynamical equation. It is clear from Eq.~(\ref{friedmann}) that,
if general relativity is correct, we must have $w\equiv p/\rho < -1/3$ for the expansion of the Universe to be speeding up at some particular epoch.
Hypothetical stress-energy components obeying this relation have been termed ``dark energy."
Current observations show that $w$ today is near $-1$. 

The only logical alternative to dark energy which can explain the observational data is a 
modification of general relativity, so that the dynamics of $a(t)$ are determined by an
equation different from Eq.~(\ref{friedmann}). A variety of attempts have been made so far in this direction
(see, e.g., \cite{def02,car05,zlo08}),
although modifying general relativity on cosmological scales while still preserving successes on
solar system scales and also matching available cosmological data on structure formation is challenging. Modified gravitation theories also tend to be more difficult to solve than
general relativity, so detailed cosmological predictions for a given theory are often lacking. 

How can we distinguish between general relativity plus dark energy and modified gravity
in a model-independent way? A number of papers have pointed out that in general relativity,
a specific relationship exists between two basic gravitational phenomena in cosmology:
the expansion history of the Universe and the growth of cosmic structure \cite{lin05,ish06,kno06,jai08}. 
For scales well inside the cosmological horizon, the linear-theory growth factor $D(a)$ 
%as a function of the cosmic scale factor $a$ 
is determined by the differential equation (see, e.g., \cite{Dodelson})
\begin{equation}
\frac{d^2D}{da^2} + \left(\frac{3}{a} + \frac{1}{H}\frac{dH}{da}\right)\frac{dD}{da} - \frac{3\Omega_m(a)H_0^2}{2a^5H^2} D=0,
\label{growth_formula}
\end{equation}
where the Hubble parameter $H(a)\equiv (1/a)(da/dt)$, $H_0$ is the present value of $H$, and
$\Omega_m(a)\equiv 8\pi G \rho_m(a)/(3H(a)^2)$ is the ratio of 
the matter density $\rho_m$ to the critical density. (This equation assumes that any energy density components besides matter and radiation have negligible density variations.) The solution to this
equation can be described to a very good approximation by 
\begin{equation}
\frac{d\ln D}{d\ln a} \simeq \Omega_m(a)^\gamma
\label{growth_index}
\end{equation}
for a wide range of realistic models \cite{lin05}. The point behind this
useful parameterization is the separation of the effect of expansion
history, encapsulated in the function $\Omega_m(a)$, from the 
growth rate, conveniently described by the single exponent $\gamma$. 
For standard cosmological models with dark energy, $\gamma \simeq 0.55 + 0.05[1+w(a=0.5)]$
(reducing to the familiar $\gamma = 0.6$ for $w=0$).
Sophisticated and general parameterizations of the evolution of the scale factor
and structure formation in theories different than general relativity have been
constructed \cite{hu07,hu08}, but Eq.~(\ref{growth_index}) provides a simple, single-parameter
relation valid for general relativity that can be observationally tested in principle. 
While a given modified gravity theory is not guaranteed to have a substantially
different value for $\gamma$ than general relativity with dark energy (see \cite{kun07} for an example), generically this will be true. 
Linder and Kahn, for example, calculates that DGP gravity \cite{dva00} gives $\gamma\simeq 0.68$ \cite{lin07}
and give $\gamma$ for various scalar-tensor theories.

Testing Eq.~(\ref{growth_index}) observationally is not an easy task, however. It requires an
observable which measures the growth rate of structure with good precision over a large range of redshifts.
Directly measuring galaxy clustering as a function of redshift can in principle give the linear
growth factor if the bias factor between galaxy clustering and the underlying mass distribution
is understood well enough and if surveys complete to high enough redshift are available. 
Linder has advocated using redshift-space distortions, which are a measure of the internal
velocity dispersions of bound objects like galaxy clusters \cite{lin08}: up to a bias factor relating
the galaxy distribution to the underlying mass distribution, the redshift space distortion is
proportional to the left side of Eq.~(\ref{growth_index}). This technique is promising, but relies on
dynamics in the nonlinear regime, and requires spectroscopic redshifts of many galaxies. Both of these
optical observation methods become increasingly difficult at high redshift.

We advocate a different approach to testing Eq.~(\ref{growth_index}): velocities of galaxy clusters
obtained from the kinematic Sunyaev-Zeldovich (kSZ) effect. The microwave background
radiation has its blackbody temperature shifted as it passes through a galaxy cluster, with the
temperature shift being proportional to the line-of-sight velocity of the cluster and to its total
optical depth for Compton scattering of the microwave radiation \cite{sz80}. For typical 
masses and velocities of large galaxy clusters, the temperature shift will be on the order of a few micro-Kelvin,
on an angular scale of around one arcminute.  The Sunyaev-Zeldovich effect is a powerful
probe of cosmology because it is essentially independent of cluster distance. Also, the kSZ effect directly
measures velocities with respect to the cosmic rest frame, unlike redshift-based velocity
measurements which generally must contend with cosmological redshifts that are much larger than the
redshift due to velocities. The kSZ effect, a temperature shift on the 
order of one part in a million of the mean background temperature, has not yet been detected
(see \cite{hol97,ben03} for upper limits), but a new generation of experiments \cite{ACT,hincks09,SPT,stan09,nor09} are now making
measurements at these angular scales and have the potential to detect the kSZ effect in clusters. 

Here we simply assume that future Sunyaev-Zeldovich surveys will result in a galaxy cluster
velocity catalog, with each cluster having its sky position and redshift known exactly and its
line-of-sight velocity determined with some characteristic error. From such a catalog, a number of
different velocity statistics can be formed which are useful for cosmology. We have previously
demonstrated the utility of cluster velocity statistics for constraining properties of dark energy
\cite{bha07,bha08a}. Consider in particular the mean pairwise cluster relative velocity \cite{dav77,jus00}, 
which can
be estimated from only observed line-of-sight velocity components \cite{bha08a}. Using a pair conservation
equation,
an analytic approximation for the mean relative velocity for two clusters separated by a 
comoving distance $r$ and at an average scale factor $a$ is \cite{sheth01}
\begin{equation}
v_{ij}(r,a) = -\frac{2}{3} aH(a)\frac{d\ln D}{d\ln a}\frac{r{\bar\xi(r,a)}}{1+\xi(r,a)},
\label{vij}
\end{equation}
where $\xi(r,a)$ is the cluster two-point correlation function and $\bar\xi(r,a)$ is the
correlation function averaged over a sphere of radius $r$. Both of these correlation functions
can be computed from the matter power spectrum and a bias giving the average number
of clusters which form in a given overdensity. 
We assume that this bias is given by the standard LCDM cosmology \cite{bha08a}, which may not
be precisely valid for alternate theories of gravity.  However, numerical simulations \cite{hu09} suggest that  deviations in the large-scale bias for alternate theories of gravity is only a few percent, so we expect 
this assumption to have little effect on our results.

Notice that the amplitude of Eq.~(\ref{vij}) is proportional to $d\ln D/d\ln a$, given by Eq.~(\ref{growth_index}).
So clearly this statistic can be used to measure the structure growth index $\gamma$.  To quantify this
assertion, we have computed the constraints on a 5-parameter standard $\Lambda$CDM cosmological
model from a 
cluster velocity catalog with a given number of clusters and a given mean velocity error, combined with
priors on each parameter expected from the Planck Satellite's upcoming measurement of the primary
microwave background temperature fluctuations \cite{detf},
and a measurement of $H_0=72\pm 8$ km/s/Mpc 
from the Hubble Key Project \cite{freedman01}. We perform a standard Fisher Matrix
estimate \cite{jungman96} of the constrained region in the multi-dimensional parameter space consisting of the
amplitude of density fluctuations $\sigma_8$, the power-law index of the primordial density perturbations $n$,
the Hubble parameter today $H_0$, the present matter density $\Omega_m$, and
the growth index $\gamma$. For simplicity we assume the Universe is spatially flat, as indicated
by current observations; the fiducial model for the scale factor evolution is standard $\Lambda$CDM.
The resulting constraint on the growth index $\gamma$ for a cluster velocity catalog
with 4000 cluster velocities, marginalized
over the other parameters, is shown in Figure 1. The horizontal axis gives the standard error for measuring 
each cluster line-of-sight velocity and the vertical axis gives $\sigma(\gamma)$, 
the 1-$\sigma$ standard error on the
resulting measurement of $\gamma$. 

\begin{figure}
\includegraphics[width=10cm]{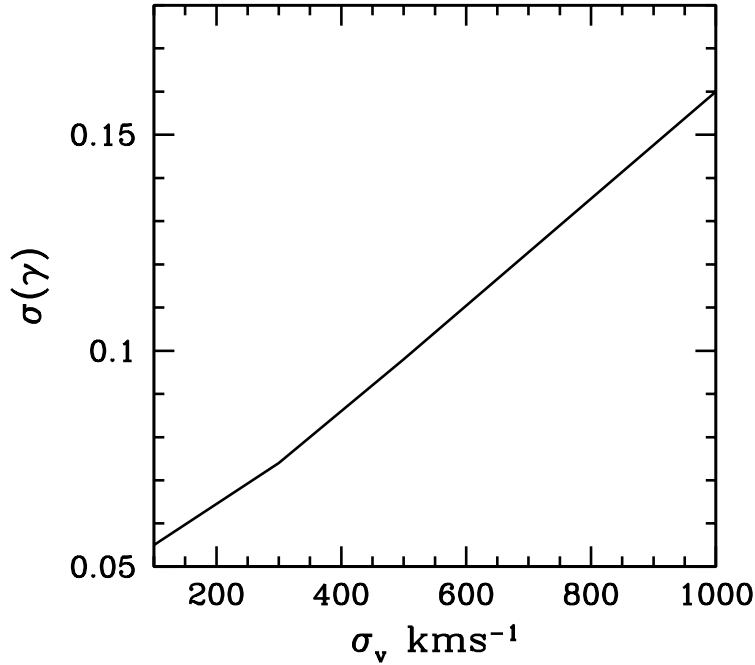}
\caption{The 1-sigma constraint on the growth index $\gamma$ in a seven-parameter spatially flat
cosmological model,
constrained by a galaxy cluster velocity survey of 4000 clusters chosen via the Sunyaev-Zeldovich effect,
as a function of assumed velocity errors. Prior parameter constraints anticipated from the Planck satellite measurement
of the primary microwave background power spectrum, plus the current Hubble Key Project constraint
on the Hubble parameter, are included.}
\end{figure}

For the current best-fit cosmological model, 4000 clusters
corresponds to measuring a velocity for all clusters with masses larger than $2\times 10^{14}\,M_\odot$
in around 400 square degrees of sky. This mass is near the anticipated cluster detection threshold for
current Sunyaev-Zeldovich experiments, although they are still a ways from having the sensitivity to
measure the smaller kinematic SZ signal. If we measure the cluster velocities with an error
of 400 km/s via their kinematic SZ distortion, this will provide a measurement of $\gamma$ to 0.08,
a level which is interesting for discriminating various modified gravity theories from dark energy.

This simple calculation likely gives a conservative estimate of $\sigma(\gamma)$,
for a number of reasons.
The error is statistics dominated, so it can be decreased by measuring velocities for clusters in
a larger sky region, with $\sigma(\gamma)$ scaling like the inverse square root of the sky area. 
Measuring velocities more precisely can also somewhat increase the
precision in measuring $\gamma$, although internal motions of cluster gas provide an astrophysical
limit of around 100 km/s to how well cluster velocities can be measured with the kSZ effect \cite{nagai03}.
The prior on cosmological parameters
used here also does not account for correlations between various parameters constrained
by the microwave background power spectrum; a more detailed parameter space investigation
will likely result in smaller errors on $\gamma$. Including additional velocity statistics or other
measures of structure formation also may decrease the
error on $\gamma$, provided correlations between the various statistics are correctly accounted for. 
Finally, including $\gamma$ as an extra parameter generally does not substantially degrade the
simultaneous constraints on other cosmological parameters \cite{hut07}.

The current experiments ACT and SPT are mapping large portions of the sky at arcminute
resolution in multiple microwave frequency bands. Such measurements will constrain
certain linear combinations of gas temperature, line-of-sight gas mass, and line-of-sight gas
velocity, depending on frequencies and noise level \cite{seh05}. The addition of a gas
temperature determination from X-ray measurements often greatly improves the precision
of cluster velocity determinations. While no cluster peculiar velocities have yet been measured,
velocity catalogs for hundreds or thousands of clusters are clearly within reach as the noise level
of microwave maps decreases. 

The kinematic SZ effect for galaxy clusters provides a unique window into the growth of structure
in the Universe. Like its thermal SZ counterpart, it is essentially independent of cluster distance,
so it can probe structure growth over all epochs and over huge volumes. But in contrast to
thermal SZ cluster detection to measure the evolution of cluster number density, cluster velocities
derived from the kinematic
SZ signal depend only weakly on cluster mass \cite{bha08b}
(since the gravitational field causing cluster peculiar velocities provides the same acceleration to 
all masses),
sidestepping systematic uncertainties related to the connection of cluster mass to observed SZ
signal. The experimental challenge is daunting: detection of the tiny blackbody kinematic SZ distortion
at arcminute resolution in multiple frequency bands, and disentangling this signal from other
larger contributions including the thermal SZ distortion, dust emission, and sub-millimeter galaxy
emission (e.g., \cite{aus09} for a recent measurement). But progress has been rapid, 
and the payoff is one of the few reliable methods
available for probing the fundamental properties of gravitation on cosmological scales, perhaps
shedding light on the accelerating expansion of the Universe.

\begin{acknowledgments}
% We thank XXX for helpful discussions.
S.B.\ was supported by a Mellon Fellowship
while at the University of Pittsburgh. 
A.K.\  gratefully acknowledges support
from NSF grant AST-0807790.
\end{acknowledgments}

%%%%%%%%%%%%%%%%%%


\begin{thebibliography}{}

\bibitem{fri08} J.~Frieman, M.S.~Turner, and D.~Huterer, Ann.\ Rev.\ Astron.\ Astrophys.\ {\bf 46}, 385 (2008). 
% Review of accelerating universe, possible solutions through dark energy or modified gravity
 
\bibitem{def02} C.~Deffayet, G.~Dvali, and G.~Gabadadze, Phys.\ Rev.\ D {\bf 65}, 044023 (2002). 
% Accelerated Universe from Gravity Leaking to Extra Dimensions 

\bibitem{car05} S.M.~Carroll et al., Phys.\ Rev.\ D {\bf 71}, 063513 (2005). 
% The Cosmology of Generalized Modified Gravity Models
% Explores general late-time properties of curvature-invariant modifications of Hilbert action
% which depart from GR in low-curvature regime. 

\bibitem{zlo08} T.G.~Zlosnik, P.G.~Ferreira, and G.D.~Starkman, Phys.\ Rev.\ D {\bf 77}, 084010 (2008).
% On the growth of structure in theories with a dynamical preferred frame
% Find Einstein-Aether solutions which have accelerated expansion at late times

\bibitem{lin05} E.V.~Linder, Phys.\ Rev.\ D {\bf 72}, 043529 (2005).
% Cosmic Growth History and Expansion History. Introduces growth index.

\bibitem{ish06} M.~Ishak, A.~Upadhye, and D.N.~Spergel, Phys.\ Rev.\ D {\bf 74}, 043513 (2006).
% Probing Cosmic Acceleration Beyond the Equation of State: Distinguishing Between Dark
% Energy and Modified Gravity Models

\bibitem{kno06} L.~Knox, Y.S.~Song, and J.A.~Tyson, Phys.\ Rev.\ D {\bf 74}, 023512 (2006). 
% comparing expansion rate and growth function using weak lensing

\bibitem{jai08} B.~Jain and P.~Zhang, Phys.\ Rev.\ D {\bf 78}, 063503 (2008).
% Observational Tests of Modified Gravity

\bibitem{Dodelson} S.~Dodelson, {\sl Modern Cosmology} (Academic Press, 2003).
% Dodelson textbook 

\bibitem{hu07} W.~Hu and I.~Sawicki, Phys.\ Rev.\ D {\bf 76}, 104043 (2007).
% A Parameterized Post-Friedmann Framework for Modified Gravity
% general parameterization of modified gravity models which accelerate the
% expansion without dark energy

\bibitem{hu08} W.~Hu, Phys.\ Rev.\ D {\bf 77}, 103524 (2008).
% Parametrized Post-Friedmann Signatures of Acceleration in the CMB
% Gives general perturbation formalism for general departures from GR,
% multiple radiation/matter/relativistic/curvature components

\bibitem{kun07} M.~Kunz and D.~Sapone, Phys.\ Rev.\ Lett.\ {\bf 98}, 121301 (2007).
% Dark Energy versus Modified Gravity. Shows that a generalized DE model matches
% cosm. observables for DGP gravity, so indistinguishable. Cite in conclusions to show
% that there is no guarantee of distinguishing. 

\bibitem{dva00} G.~Dvali, G.~Gabadadze, and M.~Porrati, Phys.\ Lett.\ {\bf B485}, 208 (2000).
% original DGP gravity paper

\bibitem{lin07} E.V.~Linder and R.N.~Cahn, Astropart.\ Phys.\ {\bf 28}, 481 (2007). 
% Parameterized Beyond-Einstein Growth: uses gravitational growth index, derives
% values for time-varying gravity, DGP, some scalar-tensor theories

\bibitem{lin08} E.V.~Linder, Astropart.\ Phys.\ {\bf 29}, 336 (2008). 
% Redshift Distortions as a Probe of Gravity

\bibitem{sz80} R.~Sunyaev and Y.B.~Zeldovich, Mon.\ Not.\ R.\ Astron.\ Soc.\ {\bf 190}, 413 (1980). 
% first kSZ paper

\bibitem{hol97} W.L.~Holzapfel et al., Astrophys.\ J.\ {\bf 481}, 35 (1997).
% First kSZ limits

\bibitem{ben03} B.A.~Benson et al., Astrophys.\ J.\ {\bf 592}, 674 (2003).
% Peculiar Velocity Limits from Measurements of the Spectrum of the Sunyaev-Zel'dovich 
% Effect in Six Clusters of Galaxies 

\bibitem{ACT} A.~Kosowsky et al., New Astron.\ Rev.\ {\bf 50}, 969 (2006).
% The Atacama Cosmology Telescope Project: A Progress Report

\bibitem{hincks09} A.~Hincks et al., Astrophys.\ J.\ submitted (2009), arXiv:0907.0461.
% First ACT paper

\bibitem{SPT} J.~Ruhl et al., Proc.\ SPIE {\bf 5498}, 11 (2004). 
% The South Pole Telescope

\bibitem{stan09} Z.~Staniszewski et al., Astrophys.\ J.\ submitted (2009), arXiv:0810.1578.
% First SPT paper

\bibitem{nor09} M.~Nord et al., Astron.\ Astrophys.\ submitted (2009), arXiv:0902.2131.
% APEX-SZ paper, SZ of Abell 2163

\bibitem{bha07} S.~Bhattacharya and A.~Kosowsky, Astrophys.\ J.\ Lett.\ {\bf 659}, 83 (2007).
% Cosmological Constraints from Galaxy Cluster Velocity Statistics

\bibitem{bha08a} S.~Bhattacharya and A.~Kosowsky, Phys.\ Rev.\ D {\bf 77}, 083004 (2008).
% Dark Energy Constraints from Galaxy Cluster Peculiar Velocities

\bibitem{dav77} M.~Davis and P.J.E.~Peebles, Astrophys.\ J.\ Suppl.\ {\bf 34}, 425 (1977).
% Pairwise peculiar velocity

\bibitem{jus00} R.~Juszkiewicz et al., Science {\bf 287}, 109 (2000). 
% Uses pairwise velocity of galaxies to constrain Omega

\bibitem{sheth01} R.K.~Sheth et al., Mon.\ Not.\ R.\ Ast.\ Soc.\ {\bf 325},
1288 (2001).
% Analytic formulas for velocity statistics

\bibitem{hu09} F.~Schmidt et al., Phys.\ Rev.\ D {\bf 79}, 083518 (2009).

\bibitem{detf} A.~Albrecht et al., arXiv:astro-ph/0609591 (2006).
% Dark Energy Task Force report

\bibitem{freedman01} W.L.~Freedman et al., Astrophys.\ J.\ {\bf 553}, 47 (2001).
% HST Key Project H0 = 72\pm 8

\bibitem{jungman96} G.~Jungman et al., Phys.\ Rev.\ D {\bf 54}, 1332 (1996).
% Refer to for original Fisher Matrix calculation

\bibitem{nagai03} D.~Nagai, A.V.~Kravtsov, and A.~Kosowsky, Astrophys.\ J.\ {\bf 587}, 524 (2003). 
% Effect of Internal Flows on Sunyaev-Zeldovich Measurements of Cluster Peculiar Velocities

\bibitem{hut07} D.~Huterer and E.V.~Linder, Phys.\ Rev.\ D {\bf 75}, 023519 (2007).
% Separating dark physics from physical darkness: Minimalist modified gravity versus dark energy.
% Evaluates cosmological constraints on parameters plus growth index, showing how minimal
% degradation in other parameters while constraints on growth index, using DETF probes.
% Similar to what this paper does for velocities

\bibitem{seh05} N.~Sehgal, A.~Kosowsky, and G.~Holder, Astrophys.\ J.\ {\bf 635}, 22 (2005). 
% Constrained Cluster Parameters from Sunyaev-Zeldovich Observations

\bibitem{bha08b} S.~Bhattacharya and A.~Kosowsky, J.\ Cosm.\ Astropart.\ Phys.\ {\bf 0808}, 030 (2008).
% Systematic Errors in Sunyaev-Zeldovich Surveys of Galaxy Cluster Velocities

\bibitem{aus09} J.E.~Austermann et al., submitted to Mon.\ Not.\ R.\ Ast.\ Soc.\ (2009), arXiv:0907.1093.
% AzTEC catalog

%\bibitem{mor09} M.J.~Mortonson, W.~Hu, and D.~Huterer, Phys.\ Rev.\ D {\bf 79}, 023004 (2009).
% Falsifying Paradigms for Cosmic Acceleration
% Systematic look at how well consistency relations can be probed for LCDM, quintessence, and
% general smooth dark energy


\end{thebibliography}
\end{document}